\documentclass{epl}
\usepackage{graphicx}

\title{Frequency-domain study of relaxation in a
       spin glass model for the structural glass transition}
    
\shorttitle{Frequency-domain study ...}
\author{F. Rao\inst{1}, A. Crisanti\inst{2} and F. Ritort\inst{3}}

\institute{
\inst{1} Department of Biochemistry, University of Zurich, 
          Winterthurerstrasse 190, CH-8057 Zurich,
                  Switzerland.\\
\inst{2} Dipartimento di Fisica, Universit\`a di Roma, ``La Sapienza'' 
         and
         INFM unit\`a di Roma I and SMC, 
         P. le A. Moro 2, 00186, Rome, Italy. \\
\inst{3} Departament de F\'{\i}sica Fonamental, Facultat de
                    F\'{\i}sica, Universitat de Barcelona, Diagonal 647, 
                    08028 Barcelona,  Spain.
}

\rec{}{}


\date{printout: 01.04.2003}
\begin{document}

\maketitle  
        
\begin{abstract}
  We have computed the time-dependent susceptibility for the
  finite-size mean-field Random Orthogonal model (ROM). We find that
  for temperatures above the mode-coupling temperature the imaginary
  part of the susceptibility $\chi''(\nu)$ obeys the scaling forms
  proposed for glass-forming liquids. Furthermore, as the temperature
  is lowered the peak frequency of $\chi''$ decreases following a
  Vogel-Fulcher law with a critical temperature remarkably close to
  the known critical temperature $T_c$ where the configurational
  entropy vanishes.
\end{abstract}

A model Hamiltonian or an effective Lagrangian capable of describing 
relaxation processes in supercooled liquids and structural glasses is difficult
to obtain. However,
starting with the work of Kirkpatrick, Thirumalai and Wolynes 
\cite{KirWol87b,KirThi87a} in the 
late 80's, it is now clear that there is a close analogy between 
some mean-field spin-glass models and structural glasses \cite{BCKM}. 
The former thus provide a set of microscopical models where glassy
dynamics can be studied analytically. 
The basic simplification occurring in mean-field models is that after 
averaging over the disorder and making the number of spin very large 
($N\to\infty$) one is left with a closed set of equations for the two-time
correlation and response functions which, above a critical temperature, are
basically equivalent to the schematic mode coupling equations introduced
by Leutheusser, G\"{o}tze and others \cite{SMCT}
as a model for the ideal glass transition.

In mean-field models the barrier separating different ergodic
components diverges in the mean-field limit, hence at the critical
temperature $T_D$ a real ergodic to non-ergodic transition takes place
with diverging relaxation times since barriers cannot be overcome, 
in agreement with mode-coupling theories (MCT) where $T_D$ coincides with the 
MCT critical temperature $T_{MCT}$. In what
follows we will denote $T_D$ by $T_{MCT}$. 
In real systems $T_{MCT}$ still denotes the separation temperature
for activated hopping dominated dynamics \cite{Hop},
however, barriers are of finite height and the glass
transition appears at $T_g < T_{MCT}$ where the typical activation time
over barriers is of the same order of the observation time.  To go
beyond mean-field it is necessary to include activated processes, a
very difficult task since it implies the knowledge of excitations
involved in the dynamics.

Recently it has been shown how activated processes in mean-field models 
could be included just keeping $N$ finite \cite{CriRit00}, and hence giving 
support to the scenario of the fragile glass transition developed
from spin-glass models. This is not
a trivial assumption since it is not {\it a priori} clear why
excitations in mean-field spin glass models should have similar properties 
to those of supercooled liquids. This, for example, seems to be the case
for the Random Orthogonal model (ROM) \cite{ROM1}, but not 
for the mean-field Potts-Glass model as indicated by recent studies of
its finite-size version \cite{Potts}. 

In a series of works the spectral properties of the primary or
$\alpha$ relaxation in supercooled liquids has been largely studied by
means of dielectric spectroscopy
\cite{Dixetal90,SchKreSch91,MenNag95,KudBenLenRos95,LehNag97}.  The
main conclusion is a there is a three-parameter scaling function that
allows to collapse the imaginary part $\epsilon''(\nu)$ of
the dielectric susceptibility $\epsilon(\nu)$ at different temperatures and
for several glass-forming liquids onto a master curve. The master plot
is able to reproduce the $\epsilon''(\nu)$ data around the relaxation
peak $\nu_p$ and also at higher frequencies. The bad collapse in the low
frequency part has been object of some debates
\cite{SchKreSch91,KudBenLenRos95}, however there is no dispute above
$\nu_p$. The frequency $\nu_p$ has a very strong temperature
dependence, commonly fitted by a Vogel-Fulcher form ${\rm
  log}_{10}(\nu_p) = {\rm log}_{10}(\nu_0) - A/(T-T_0)$, where $T_0$
is close to the Kauzmann temperature \cite{Kauzmann48} where the
configurational entropy vanishes \cite{Angell88}.

In this contribution we compare the frequency-domain analysis of the
finite-size ROM with the above scenario. The main motivation for this
study was to make a stringent test on the ROM as a possible model for
the fragile-glass scenario.  The range of temperatures we explore in
this paper are all above the mode-coupling transition $T_{MCT}$,
therefore we do not expect to find diverging timescales in the large
$N$ limit. In the light of the experimental
results~\cite{Dixetal90,SchKreSch91,MenNag95,KudBenLenRos95,LehNag97},
that also explore the region above $T_{MCT}$, we have
decided to focus our attention in this regime in the framework
of finite-size mean-field models. Here, we do not address the behavior
below $T_{MCT}$ where a $N$ dependence of the frequency response is
expected.

The ROM \cite{ROM1} is defined by the Hamiltonian
\begin{equation}
\label{eq:ham}
  H = - 2 \sum_{ij} J_{ij}\, \sigma_i\, \sigma_j 
      - h\sum_{i}\, \sigma_i
\end{equation}
where $\sigma_i=\pm 1$ are $N$ Ising spin variables, and $J_{ij}$ is a
$N\times N$ random symmetric orthogonal matrix with $J_{ii}=0$.  This
model was proposed in \cite{ROM1} as a random extension of some
non-disordered models describing low autocorrelation binary sequences.
Numerical simulations are performed using the Monte Carlo method with
the Glauber algorithm for temperatures in the range $0.6$ up to $2.0$.
For $N\to\infty$ and $h=0$ this model has a dynamical transition at
$T_{MCT}=0.536$, and a static transition at $T_c=0.256...$ \cite{ROM1}.  To
study the frequency response we considered a time-dependent field of
the form $h(t) = h_0\cos(2\,\pi\,\nu t)$, where the time is measured
in MC steps and $h_0=0.2$ small enough to be within
the linear response regime. In our simulations the typical range of
$\nu$ was $10^{-6}$ -- $10^{-1}$. For each frequency we have measured
the complex susceptibility $\chi(\nu) = \chi'(\nu) + i\chi''(\nu)$ as
\begin{equation}
\label{eq:chip}
\chi'(\nu) = \frac{1}{NM} \sum_{t=1}^{M}\sum_{j=1}^{N} \sigma_j(t)\, 
                               \cos(2\,\pi\,\nu t),
\end{equation}
\begin{equation}
\label{eq:chipp}
\chi''(\nu) = \frac{1}{NM} \sum_{t=1}^{M}\sum_{j=1}^{N} \sigma_j(t)\, 
                               \sin(2\,\pi\,\nu t).
\end{equation}
The number of MC steps $M$ after equilibration was $100$ for the
largest $\nu$ and up to $10^7$ for the shortest $\nu$. We show here
results for $N=300$, the finite-size dependence will be discussed
below. The choice of $N=300$ is a good compromise between small
sample-to-sample fluctuations and small barriers height.

\begin{figure}[hbt!]
  \centering
  \includegraphics[scale=0.9]{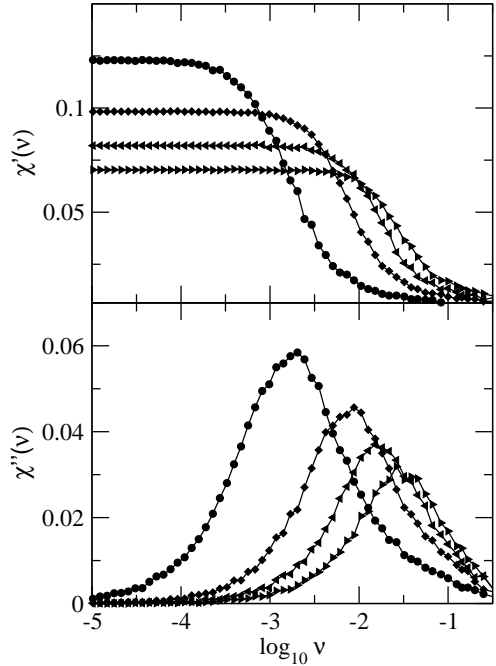}
  \caption{The real and imaginary part of the complex susceptibility 
    $\chi'$ and $\chi''$ as function of $\nu$ for the ROM with $N=300$ at
    different temperatures. Upper panel top to bottom, lower panel
    left to right, $T=0.7$, $0.9$, $1.1$ and $1.3$.
  }
  \label{fig:suc300}
\end{figure}
Figure  \ref{fig:suc300} shows the real and imaginary parts of the 
susceptibility over the available range of frequency. Not all temperatures
are reported for a better drawing.
The relaxation peak in the imaginary part can be fitted with a log-normal
form \cite{WuNag92}:
\begin{equation}
 \label{eq:logn}
 \chi''(\nu) = \frac{\Delta\chi}{\sqrt{\pi}\Sigma}
          \exp\left[-({\rm log}_{10}\nu - {\rm log}_{10}\nu_p)^2/\Sigma^2
               \right]
\end{equation}
where $\nu_p$ is the frequency of the peak, $\Sigma$ the width,
and $\Delta\chi = \chi'_0 - \chi'_{\infty}$, where $\chi'_0$ and
$\chi'_\infty$ are, respectively, 
the low and high frequency limit of $\chi'(\nu)$.

\begin{figure}[hbt!]
  \centering
  \includegraphics[scale=0.9]{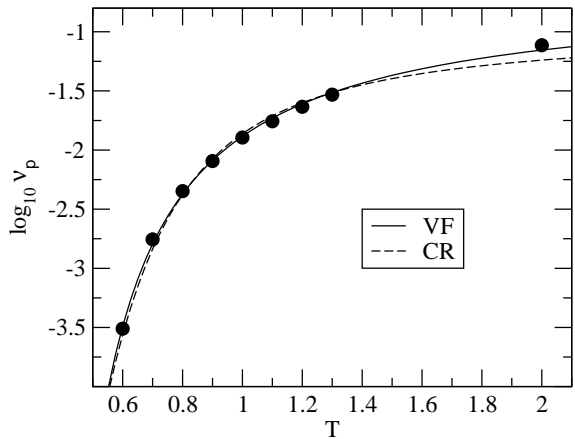}
  \caption{${\rm log}_{10}\nu_p$ as function of $T$ for the ROM with
  $N=300$. 
   The full line is the Vogel-Fulcher
    law $\nu_p = \nu_0 \exp[ -A /(T-T_0) ]$ while
    the dashed line is the formula
  $\nu_p = \nu_0 \exp[ -A \beta_{\rm eff}(T)/T ]$ of Ref. 
  \cite{CriRit02} with 
  $\beta_{\rm eff}(T) = \partial S_c(e_{\rm is}) / \partial e_{\rm is} 
                     |_{e_{\rm is}=e_{\rm is}(T)}$ evaluated using the
 results of Refs. \cite{CriRit00}.
  }
  \label{fig:vogel-fit300}
\end{figure}

As the temperature is lowered the peak 
frequency $\nu_p$ decreases, and the width $\Sigma$ broadens.  
The behavior of $\nu_p$ is consistent with the Vogel-Fulcher law
$\exp[-A/(T-T_0)]$ \cite{MenNag95}. 
The fit of the frequency peak $\nu_p$ for the ROM 
with the Vogel-Fulcher formula is rather good,
see Fig. \ref{fig:vogel-fit300},
and gives
$A = 0.89\pm 0.06$ and $\ln\nu_0 = 0.64\pm 0.02$
$T_0= 0.28\pm 0.02$, a value in agreement with 
the critical value $T_c = 0.256...$. We note, however, that data can also be
fitted using different expressions such as $\exp[-A/T^2]$ or the 
Adam-Gibbs formula $\exp[-A/(TS_{c}(T))]$ where $S_{c}$ is the
configurational entropy. 
In particular we note that the data can be also fitted with the 
formula (CR)
$\nu_p = \nu_0 \exp[ -A \beta_{\rm eff}(T)/T ]$ 
where 
$\beta_{\rm eff}(T) = \partial S_c(e_{\rm is}) / \partial e_{\rm is} 
                     |_{e_{\rm is}=e_{\rm is}(T)}$ 
derived from a cooperative scenario of relaxation
\cite{CriRit02}, see Fig. \ref{fig:vogel-fit300}. The discrepancy at high
temperature is probably due to a poor numerical estimation of the 
configurational entropy, indeed a similar deviation is found using the
Adam-Gibbs formula (not reported). We note that the CR formula predicts
a crossover form fragile to strong behaviors as the temperature is lowered.
However differences among all the above expressions can be 
appreciated only for very low values of $\nu_p$ which are out of
our measurements range. 

\begin{figure}[hbt!]
  \centering
  \includegraphics[scale=0.85]{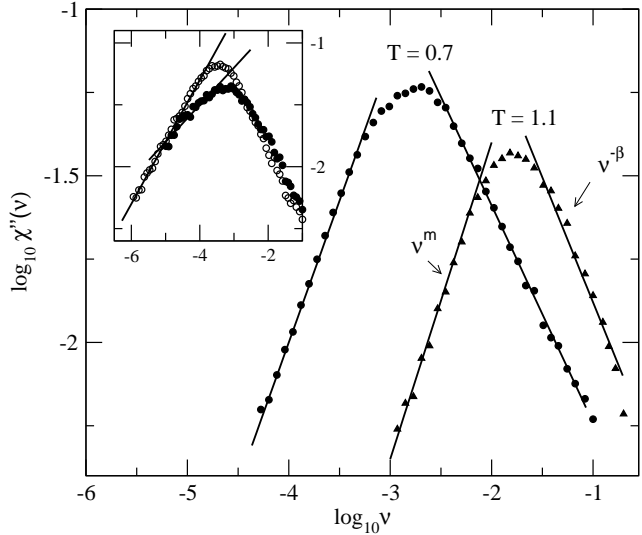}
  \caption{$\chi''(\nu)$ for the ROM with $N=300$ for
    temperatures $T=0.7$ and $1.1$. The exponents are $m=1$, $\beta=1.$ for 
    $T=1.1$ and $m=.96$, $\beta=.71$ for $T=0.7$.
    Inset: $\chi''(\nu)$ at temperature $T=0.6$ for $N=64$ (filled circles) 
    and $300$ (empty circles). The lines have slope $m=0.3$ for $N=64$
    and $m=0.5$ for $N=300$. The increase of $m$ toward $1$ as
    $N$ grows is clearly seen.
  }
\label{fig:betafit300}
\end{figure}

The analysis of the response for glass-former liquids reveals three power laws
for $\chi''$ \cite{LehNag97}: 
 $\chi''\sim \nu^{m}$ for $\nu<\nu_p$, $\chi''\sim \nu^{-\beta}$ just above
$\nu_p$ and  $\chi''\sim \nu^{-\sigma}$ for larger values $\nu\gg \nu_p$. 
The use of Monte Carlo dynamics prevents us from resolving the last
behavior due to the discreteness of the time step,
nevertheless the first two regimes are clearly seen, as indicated in Figure
\ref{fig:betafit300}. At higher temperatures $m=\beta=1$ and the
relaxation is Debye-like. As the temperature is lowered the value of $\beta$ 
decreases below $1$.
%
%
It is known that for glass-former liquids  $\beta$ and $\sigma$  
are related by $(\sigma+1)/(\beta+1) = \gamma$, where $\gamma$ is a
constant \cite{LehNag97}. 
Furthermore $\sigma$ varies linearly with temperature:
$\sigma = B(T-T_\sigma)$ with $T_\sigma\simeq T_0$ \cite{LehNag97}, 
This implies that 
$\beta = B'(T-T_0) + (1-\gamma)/\gamma$.
Inserting into this formula the values of $\beta$ obtained for the ROM 
at various temperatures and the value of $T_0$ computed from $\nu_p$ we find
$\gamma = 0.72\pm 0.02$ the same value 
found for real liquids \cite{MenNag95,LehNag97}.  

Analyticity of $\chi(\nu)$ and linearity of absorption at 
asymptotically low frequencies implies that $\chi''(\nu) \propto \nu$ for
$\nu\ll\nu_p$ \cite{SchKreSch91}. 
For the ROM with $N=300$ we find
$m\simeq 1$ for temperatures down to about $0.8$. Below $T=0.8$
significant deviations with $m<1$ are observed.  The temperature where
$m=1$ breaks down decreases as $N$ increases.
Similar deviations have been observed in data from glass-forming liquids and
generated some controversy \cite{Dixetal90,SchKreSch91,KudBenLenRos95}
on the reliability of the scaling form proposed by Dixon {\it et al.} 
\cite{Dixetal90}. Many liquids also posses secondary relaxations which overlap
the primary response broadening the peak and leading to deviation from 
linearity \cite{KudBenLenRos95}. 
In the case of ROM these secondary relaxations
are related to the fact that the barriers separating 
the low states sampled as the temperature
is decreased toward $T_{MCT}$ are not well separated for not too large $N$.
Indeed studies of mean-field spin-glass models for the structural glass 
transition shows that in the thermodynamic limit 
there is no gap between saddles separating 
local minima with energy above the threshold energy associated with 
the dynamical transition \cite{CavGiaPar97}.
This is a situation more reminiscent of spin-glasses rather than glasses.
Indeed both experimental \cite{BitMenNagRosAep96} and numerical 
simulations \cite{BitMenNagRosAep96,Rao01} show a broader shape
of $\chi''(\nu)$ near the peak.

This scenario is supported by a finite-size scaling analysis of the ROM.
Indeed we find that for a fixed temperature 
while $\beta$ is independent on $N$, the
value of $m$, when less than $1$, increases toward $1$ as $N$ is increased.
In the inset of 
Figure \ref{fig:betafit300} we show $\chi''(\nu)$ at $T=0.6$ for 
systems of size $N=64$ and $N=300$. The increase of $m$ going from $64$ to 
$300$ spins is well evident. 
%
%
We now finally address the goodness of the Nagel scaling.
Dixon {\it et al} \cite{Dixetal90} have shown that all data for the dielectric 
susceptibility $\epsilon(\nu)$ of different glass-forming liquids and temperatures
can be collapsed onto a single master curve by plotting
$(1/w){\rm log}_{10}(\epsilon''(\nu)\nu_p/\nu\Delta\epsilon)$ versus
$(1/w)(1/w+1){\rm log}_{10}(\nu/\nu_p)$, where 
$\Delta\epsilon= \epsilon'(0)-\epsilon'(\infty)$ 
is the relaxation strength measured by the step height of the real
part of $\epsilon(\nu)$, $\nu_p$ is the peak frequency, and $w$ is the 
half-maximum width of the $\epsilon''(\nu)$ peak normalized with the 
corresponding width of the Debye peak: $w_{D} \simeq 1.14$ decades. 

\begin{figure}[hbt!]
  \centering
  \includegraphics[scale=0.9]{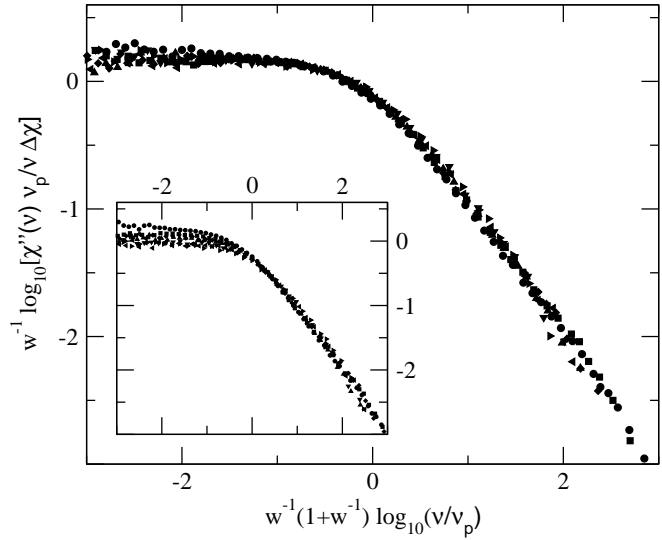}
\caption{The Nagel's scaling for the ROM with $N=300$ and temperatures
  $T=2.0$, $1.3$, $1.2$, $1.1$, $1.0$, $0.9$, $0.8$, $0.7$ and $0.6$.
  See text for details.
        }
\label{fig:scaling300}
\end{figure}

Figure \ref{fig:scaling300} shows the Nagel plot for the ROM with $N=300$. The data 
are from temperatures ranging from $0.6$ up to $2.0$.  For each curve the parameters
$\Delta\chi$, $w = 2\sqrt{2}\Sigma/w_d$ and $\nu_p$ have been obtained
from the log-normal fit of $\chi''$ using (\ref{eq:logn}).
We see that while the collapse  for $\nu>\nu_p$ is good for all temperatures,
for $\nu<\nu_p$ only data with $m=1$ do collapse.
As noted in Refs. \cite{KudBenLenRos95} the optimization of the
three parameters through the fitting is essential to have a good collapse
of data. For comparison in the inset we report the Nagel plot for the
same data done using the parameters read directly from the plot of Figure
\ref{fig:suc300} without any fitting. The improvement through the fitting is 
evident.

In conclusion, we have shown that the primary relaxation in the finite
size mean-field ROM obeys the scaling form typical of glass-forming
liquids. Furthermore, the frequency peak of the imaginary part of the
complex susceptibility follows the Vogel-Fulcher law with critical
temperature $T_0=0.28\pm 0.02$ very close to the critical temperature
$T_c=0.256...$, the Kauzmann temperature of the model. All system sizes
studied (up to $N=300$) lead to this value for $T_0$. Because we used
Monte Carlo dynamics there is a maximum value for the frequency $\sim
N$ determined by the discreteness of the elementary time step. As a
consequence we are not able to resolve the second scaling behaviors of
the imaginary part of the susceptibility $\chi''(\nu)\sim
\nu^{-\sigma}$ for $\nu\gg\nu_p$.  Nevertheless by assuming that the
exponent $\sigma$ vanishes linearly at $T_0$, we obtained for the
constant $\gamma$ relating the exponents $\beta$ and $\sigma$ the
value $\gamma=0.72\pm 0.02$, the same found for real glass-forming
liquids \cite{MenNag95,LehNag97}.  Again this value does not depend on
the system size for the sizes we studied.  Changing $N$ only 
modifies the temperature below which secondary relaxations related
to barrier crossing breaks the Nagel scaling.
Overall, the present
results show that finite-size mean-field spin glasses capture the
cooperative effects responsible for the relaxational processes
observed in glass forming liquids when approaching the mode-coupling
temperature. Moreover, to our knowledge, this is the first time that
finite-size mean-field spin glasses give a quantitative prediction in
agreement with experimental data. The extension of this analysis to
the region below $T_{MCT}$ where strong finite $N$ effects are to be
observed remains an interesting open problem.

\acknowledgments
We acknowledge 
F. Sciortino and P.Tartaglia for a critical reading of the manuscript.
A.C. acknowledges support from the INFM-SMC center. F.R has been
supported by project the Spanish Ministerio de Ciencia y Tecnolog\'{\i}a
Grant number BFM2001-3525 and Generalitat de Catalunya. A.C and F.R have
also benefited from the Acciones Integradas Espa\~na-Italia HI2000-0087.

\end{document}